\documentclass[twocolumn,showpacs,aps,prl,superscriptaddress,letterpaper]{revtex4}
\usepackage{graphicx}
\usepackage{dcolumn}
\usepackage{amsmath}
\usepackage{epsfig}
\usepackage{multirow}
\usepackage{subfigure}
\usepackage{amssymb}

\long\def\inst#1{\par\nobreak\kern 4pt\nobreak
    {\itshape #1}\par\vskip 10pt plus 3pt minus 3pt}
\RequirePackage{xspace}
\usepackage{relsize}

\newcommand{\BABARPubYear}    {17}
\newcommand{\BABARPubNumber}  {001}

\newcommand{\SLACPubNumber} {16923}

\input babarsym

\def\Lumi      {$53$~fb$^{-1}$}
\def\LumiFourS {$5.9$~fb$^{-1}$}
\def\LumiThreeSOn  {$28.5$~fb$^{-1}$}
\def\LumiThreeSOff {$2.7$~fb$^{-1}$}
\def\LumiTwoSOn  {$14.4$~fb$^{-1}$}
\def\LumiTwoSOff {$1.5$~fb$^{-1}$}
\def\LumiThreeSOnLE{$20$~fb$^{-1}$}
\def\LumiHE{$53$~fb$^{-1}$}
\def\LumiLE{$35.9$~fb$^{-1}$}

\def\dm          {\ensuremath{\chi}\xspace}
\def\dmbar       {\ensuremath{{\overline\chi}}\xspace}

\def\gamgam      {\ensuremath{e^+e^-\to\gamma\gamma}\xspace}
\def\bhabha      {\ensuremath{e^+e^-\to e^+e^-\gamma}\xspace}

\def\beq{\begin{equation}}
\def\eeq{\end{equation}}
\def\bea{\begin{eqnarray}}
\def\eea{\end{eqnarray}}
\def\bq{\begin{quote}}
\def\eq{\end{quote}}
\def\bi{\begin{itemize}}
\def\ei{\end{itemize}}
\def\bc{\begin{center}}
\def\ec{\end{center}}

\newcommand{\ie}{{\em i.e.}}

\def\etal{{\em et al.}}

\begin{document}

\onecolumngrid
\hbox to \hsize{
\vbox{
\begin{flushleft}
\babar-PUB-\BABARPubYear/\BABARPubNumber \\
SLAC-PUB-\SLACPubNumber \\
\end{flushleft}
\vspace{\baselineskip}
}
\hfill
}
\vspace{-\baselineskip}
\twocolumngrid
\title{
\large \bfseries \boldmath
Search for Invisible Decays of a Dark Photon Produced in $e^+e^-$
Collisions at \babar
}

%

%
\author{J.~P.~Lees}
\author{V.~Poireau}
\author{V.~Tisserand}
\affiliation{Laboratoire d'Annecy-le-Vieux de Physique des Particules (LAPP), Universit\'e de Savoie, CNRS/IN2P3,  F-74941 Annecy-Le-Vieux, France}
\author{E.~Grauges}
\affiliation{Universitat de Barcelona, Facultat de Fisica, Departament ECM, E-08028 Barcelona, Spain }
\author{A.~Palano}
\affiliation{INFN Sezione di Bari and Dipartimento di Fisica, Universit\`a di Bari, I-70126 Bari, Italy }
\author{G.~Eigen}
\affiliation{University of Bergen, Institute of Physics, N-5007 Bergen, Norway }
\author{D.~N.~Brown}
\author{M.~Derdzinski}
\author{A.~Giuffrida}
\author{Yu.~G.~Kolomensky}
\affiliation{Lawrence Berkeley National Laboratory and University of California, Berkeley, California 94720, USA }
% ...
\collaboration{The \babar\ Collaboration}
\noaffiliation

\date{September 28, 2017}


\begin{abstract}
We search for single-photon events in \Lumi\ of \epem\ collision data
collected with the \babar\ detector at the
\pep2\ B-factory. We look for events with a single 
high-energy photon and a large missing momentum and energy, consistent
with production of a spin-1 particle $A'$ through the process
$e^+e^-\to\gamma A';\ A'\to\mathrm{invisible}$.  
Such particles, referred to as ``dark photons'', are motivated by
theories applying a $U(1)$ gauge symmetry to dark matter. 
We find no evidence for such processes and set 90\% confidence level
upper limits 
on the coupling strength of $A'$ to $e^+e^-$ in the mass range
$m_{A'}\le8$~GeV. In particular, our limits exclude the values of the
$A'$ coupling suggested by the dark-photon interpretation of the 
muon $(g-2)_\mu$ anomaly, as well as a broad range of parameters for the
dark-sector models.  
\end{abstract}


\maketitle

%

The nature of dark matter is one of the greatest mysteries of modern
physics.
It is transparent to electromagnetic radiation and we have
only been able to infer its existence through gravitational
effects. Since terrestrial searches for dark matter
interactions have so far yielded null results, it is
postulated to interact very weakly with ordinary matter. Recently,
models attempting to explain certain astrophysical
observations~\cite{ref:INTEGRAL,ref:PAMELA, ref:FERMI,Berezhiani:2013dea} as 
well as the muon $(g-2)_\mu$ anomaly~\cite{ref:g-2} have introduced an
appealing idea of a low-mass spin-1 particle, referred to as $A'$ or
$U$,
that would possess a gauge coupling of electroweak strength to
dark matter, but with a much smaller coupling to
the Standard Model (SM)
hypercharge~\cite{ref:Aprimerefs,Essig:2013lka}. Such a boson may be 
associated with a 
$U(1)$ gauge symmetry in the dark sector and kinetically mix with the
SM photon with a mixing strength $\varepsilon\ll 1$; hence the name
``dark photon''. Values as high as $\varepsilon\sim10^{-3}$ and masses in 
a GeV range have been 
predicted in the literature~\cite{ref:Aprimerefs,Essig:2013lka}. 

The decay modes of the dark photon depend on its
mass and couplings, as well as on the particle spectrum of
the dark sector. If the lowest-mass dark matter state $\dm$
is sufficiently light: $m_{\chi}<m_{A'}/2$, then the dominant decay mode of
the $A'$ is invisible: $A'\to\dm\dmbar$. The
cleanest collider signature of such particles is the production of 
monochromatic single photons in $e^+e^-\to\gamma A'$,
accompanied by significant missing energy and momentum. 
The photon energy $E_\gamma^{*}$ in the \epem center-of-mass
(CM) is related to the
missing mass  $M_X$ through $M_X^2 = s - 2E_\gamma^{*}\sqrt{s}$, where
$s$ is the square of the CM energy, and the asterisk hereafter denotes
a CM quantity. We seek a signal of the dark photon $A'$ as a narrow 
peak in the distribution of $M_X^2$ in events with a single high-energy
photon.  
As expected for the dark matter coupling
$\alpha_D<1$~\cite{Essig:2013lka}, we assume that the decay width of
the $A'$ is negligible compared to the experimental
resolution, and that the $A'$ decays predominantly to dark matter
(\ie\ the invisible branching fraction is $\approx100\%$).
Furthermore, we assume that a single $A'$ state  
exists in the range $0<m_{A'}\le8$~GeV; or if two or more states are
present, they do not interfere.

The current best limits on the mixing strength $\varepsilon$ of the dark
photon are from
searches for narrow peaks in the $\epem$ or $\mu^+\mu^-$ invariant
mass spectra~\cite{BaBarDM,KLOE,NA48,WASA,HADES,A1,APEX} and from beam-dump
and neutrino experiments~\cite{Blumlein:2013cua,Andreas:2012mt}. These limits
assume that the dominant decays of the $A'$ are to the visible SM particles,
but are not valid if there are low-mass invisible degrees of freedom.  
There are  constraints on invisible decays of the $A'$ from kaon
decays~\cite{Pospelov,E787,E949} and from the recent search for
missing energy events in electron-nucleus scattering~\cite{NA64}.

We search for the process $\epem\to\gamma A'$, followed by invisible
decays of the $A'$ in a \Lumi\ dataset~\cite{Lumi} collected with the
\babar\ detector 
at the \pep2\ asymmetric-energy \epem\ collider at the
SLAC National Accelerator Laboratory. 
The data were
collected in 2007--2008 with CM energies near the 
$\Upsilon(2S)$, $\Upsilon(3S)$, and $\Upsilon(4S)$ resonances with a
special ``single photon'' trigger described below. 
The \epem\ CM frame was boosted relative to the
detector approximately along the detector's magnetic field axis by
$\beta_z\approx0.5$.  
Since the
production of the $A'$ is not expected to be enhanced by the presence
of the $\Upsilon$ resonances, we combine the datasets collected
in the vicinity of each $\Upsilon$ resonance. 
In order to properly account for
acceptance effects and changes in the cross section as a function of
$\sqrt{s}$, we measure the signal event yields separately for the \Y2S,
\Y3S, and \Y4S\ datasets.

Since the \babar\ detector is described in detail
elsewhere~\cite{detector}, 
only the components of the detector crucial to this analysis are
 summarized below. 
Charged particle tracking is provided by a five-layer double-sided silicon
vertex tracker (SVT) and a 40-layer drift chamber (DCH). 
Photons and neutral pions are identified and measured using
the electromagnetic calorimeter (EMC), which comprises 6580 thallium-doped CsI
crystals. These systems are mounted inside a 1.5~T solenoidal
superconducting magnet. 
The Instrumented Flux Return (IFR) forms the return yoke of
the superconducting coil, instrumented in the central barrel region with
limited streamer tubes for the identification 
of muons and the detection of clusters produced
by neutral hadrons. 
We use the {\sc Geant4}~\cite{geant} software to simulate interactions
of particles 
traversing the \babar\ detector, taking into account the varying
detector conditions and beam backgrounds.

Detection of low-multiplicity single photon events requires dedicated
trigger lines. Event processing and selection proceeds in three
steps. First, the hardware-based Level-1 (L1) trigger accepts
single-photon events if they contain at least one EMC cluster with
energy above $800$~MeV (in the laboratory frame).  Second, L1-accepted
events are forwarded to a software-based Level-3 (L3) trigger, which
forms DCH tracks and EMC clusters and makes decisions for a variety of
physics signatures. Two single-photon L3 trigger lines were active
during the data-taking period. The high-energy photon line (low $M_X$,
``LowM'' hereafter) requires an isolated EMC cluster with energy
$E^{*}_\gamma>2$~GeV, and no tracks originating from the $e^+e^-$
interaction region (IR).  The ``LowM'' dataset amounts to
\LumiFourS\ collected at the \Y4S\ resonance ($\sqrt{s}=10.58$~GeV),
\LumiThreeSOn\ collected at the \Y3S\ resonance
($\sqrt{s}=10.36$~GeV), \LumiThreeSOff\ collected 30 MeV below the
\Y3S, \LumiTwoSOn\ collected at the \Y2S\ resonance
($\sqrt{s}=10.02$~GeV), and \LumiTwoSOff\ collected 30 MeV below the
\Y2S\ resonance. The total data sample collected with the ``LowM''
triggers is \LumiHE.

A low-energy 
(high $M_X$, ``HighM'') L3 single-photon trigger, which requires an EMC
cluster with  
energy $E^{*}_\gamma>1$~GeV and no tracks originating 
from the $e^+e^-$ interaction region, was active for a subset of the
data: \LumiThreeSOnLE\ collected at the
\Y3S\ resonance as well as all of the data collected below \Y3S\ and
at the \Y2S\ resonance. 
The total data
sample collected with the ``HighM'' triggers is \LumiLE.

Additional offline software filters are applied to the stored data. We
accept single-photon events if they sa\-tisfy 
one of the two following criteria. The ``LowM'' selection requires one
EMC cluster 
in the event with $E^{*}_\gamma>3$~GeV and no DCH tracks
with momentum $p^{*}>1$~GeV. The ``HighM'' selection requires 
one EMC cluster with the transverse profile consistent with an
electromagnetic shower and $E^{*}_\gamma>1.5$~GeV, and no
DCH tracks with momentum $p^{*}>0.1$~GeV. The two selection criteria
are not mutually exclusive. 

The trigger and reconstruction selections naturally split the dataset
into two broad $M_X$ ranges. The ``LowM'' selections are used for the
low $M_X$ region $-4<M_X^2<36~\mathrm{GeV}^2$. The
backgrounds in this region are 
dominated by the QED process $e^+e^-\to\gamma\gamma$, especially near 
$M_X\approx0$ ($E^{*}_\gamma\approx \sqrt{s}/2$). Due to the orientation
of the EMC crystals, which point towards the IR, one of the photons
may escape detection even if it is within the nominal EMC acceptance. 
The event selection is optimized to reduce this peaking
background as much as possible. 
The ``HighM'' trigger selection defines the high $M_X$ range
$24<M_X^2<69\ (63.5)~\mathrm{GeV}^2$ for the \Y3S\ (\Y2S)
dataset. This region is dominated by the low-angle radiative Bhabha
events $e^+e^-\to e^+e^-\gamma$, in which both the electron and the
positron escape the detector.

We suppress the SM backgrounds, which involve one or more particles
that escape detection, by requiring that a candidate event be
consistent with a single isolated photon shower in the EMC. We accept
photons in 
the polar angle range $|\cos\theta^{*}_\gamma|<0.6$, rejecting radiative
Bhabha events that strongly peak in the forward and backward
directions, and we require that the event contain no
charged particle tracks.

The signal events are further selected by a multivariate  
Boosted Decision Tree (BDT) discriminant~\cite{TMVA}, based on the
following 12 discriminating variables. 
First, after a relatively coarse selection, we include the EMC
variables that describe the shape of the electromagnetic shower: the
difference between the number of crystals in the EMC cluster and the
expectation for a single photon of given energy, and two transverse
shower moments~\cite{ref:LAT}. 
Second, we include
both the total excess EMC energy in the
laboratory frame not associated with the highest-energy photon,
and the CM energy and polar angle of the second most energetic
EMC cluster.
We also compute the azimuthal 
angle difference 
$\Delta\phi_{12}$ between the highest and second-highest energy
EMC clusters; the \gamgam\ events with partial energy deposit in the EMC
tend to peak at $\Delta\phi_{12}\sim \pi$.  
Third, a number of variables improve containment of the background
events. 
We extrapolate the missing momentum vector
to the EMC face, and compute the distance (in $(\theta,\phi)$ polar
lab-frame coordinates) to the nearest crystal edge. This allows us to
suppress $\epem\to\gamma\gamma$ events where one of the photons
penetrates the EMC between crystals leaving little detectable
energy. Furthermore, we look for energy deposited in the IFR, and compute
the correlation angle $\Delta\phi_{NH}$ between the primary photon
and the IFR cluster closest to the missing momentum
direction; \gamgam\ events produce a peak at
$\cos\Delta\phi_{NH}\sim-1$.
We also apply a fiducial selection to the azimuthal angle
$\phi_\mathrm{miss}$ of the 
missing momentum by including $\cos(6\phi_\mathrm{miss})$ into the BDT. This  
accounts for uninstrumented regions between six IFR
sectors~\cite{detector}. Finally, $\cos\theta^{*}_\gamma$ is included
in the BDT to take advantage of the different angular distributions
for signal and background events. 

The BDT discriminants are trained separately in ``LowM'' and ``HighM''
regions. Each BDT is trained using
$2.5\times10^4$ simulated signal events with uniformly distributed $A'$
masses, and $2.5\times10^4$ background events from the \Y3S\ on-peak
sample that corresponds to approximately 3~\invfb. We test the BDT,
define the final selection, and measure the signal efficiency using
sets of $2.5\times10^4$ signal and background events statistically
independent from the BDT training samples. The BDT score is designed
so that the signal peaks near 1 while the background events
are generally distributed between $-1<\mathrm{BDT}<0$. 

The event selection is optimized to minimize the expected upper limit
on the $e^+e^-\to\gamma A'$ cross section $\sigma_{A'}$.
Since the number of peaking
$e^+e^-\to\gamma\gamma$ events cannot be reliably estimated and has to
be determined from the fit to the data, this background 
limits the sensitivity to
$e^+e^-\to\gamma A'$ at the low $A'$ masses where the photon energies for
the two types of events are indistinguishable.
In this regime, we define a ``tight''
selection region $\mathcal{R}_T$ which maximizes the ratio
$\varepsilon_S/N_B$ for large $N_B$, and $\varepsilon_S/2.3$ in the limit
$N_B\to0$, where $\varepsilon_S$ is the 
selection efficiency for the signal and $N_B$ is the number of
background events expected in the full data sample.
We also
require $-0.4<\cos\theta^{*}_\gamma<0.6$ in order to 
suppress $e^+e^-\to\gamma\gamma$ events in which one of the photons would have
missed the central region of the EMC. 

A ``loose'' 
selection region $\mathcal{R}_L$ maximizes
$\varepsilon_S/\sqrt{N_B}$. This selection is appropriate at
higher $M_X$ where the background is well described by a featureless
continuum distribution, and maximal $\varepsilon_S/\sqrt{N_B}$
corresponds to the lowest upper limit on the $e^+e^-\to\gamma A'$
cross section. 

Finally, a background region $\mathcal{R}_B$
is defined by $-0.5<\mathrm{BDT}<0$ and is used to determine the $M_X^2$
distribution of the background events. The selection
criteria used in this analysis and the numbers of events selected in
different datasets are summarized in Table~\ref{tab:DS}.
\begin{table}[tb]
\caption{Datasets and event selections used in this paper. The
  characteristic energies of each dataset are listed in rows; the 
  event selections described in the text in columns. The table entries
  list the integrated luminosity and the numbers of events selected by
  each dataset. }  
\label{tab:DS}
\begin{tabular}{l|cccc|ccc}
  \hline\hline
Dataset  & \multicolumn{4}{c|}{``lowM''} & \multicolumn{3}{c}{``highM''} \\
\hline
Dataset  & $\mathcal{L}$ & \multicolumn{3}{c|}{Selection} & $\mathcal{L}$ & \multicolumn{2}{c}{Selection} \\
         &               & $\mathcal{R}_B$ & $\mathcal{R}_L^{'}$ & $\mathcal{R}_T$
& & $\mathcal{R}_B$ & $\mathcal{R}_L$ \\
\Y2S\ & $15.9\invfb$ & $22,590$ &  $\phantom{1}42$ &  $\phantom{1}6$ &
$15.9\invfb$ & $405,441$ & $324$ \\
\Y3S\ & $31.2\invfb$ & $68,476$ & $129$ & $26$ &
$22.3\invfb$ & $719,623$ & $696$ \\
\Y4S\ & $\phantom{1}5.9\invfb$ &  $\phantom{1}7,893$ &  $\phantom{1}16$ &  $\phantom{1}9$ & &         & \\
\hline
\end{tabular}
\end{table}

We measure the cross section $\sigma_{A'}$ as a function of the
assumed mass $m_{A'}$ by performing a series of unbinned extended
maximum likelihood fits to the distribution of $M_X^2$. For each value
of $m_{A'}$, varied from 0 to $8.0$~GeV in 166 steps roughly equal to half
of the mass resolution, we perform a set of simultaneous fits to \Y2S,
\Y3S, and for the low-$M_X$ region, \Y4S\ datasets. Moreover, we
subdivide the data into broad event selection bins: $\mathcal{R}_B$
used to define the background probability density functions (PDFs),
and signal regions $\mathcal{R}_L$ (used for $5.5<m_{A'}\le 8.0$~GeV),
$\mathcal{R}_T$, and 
$\mathcal{R}_L^{'}$ (used for $m_{A'}\le5.5$~GeV).
The region $\mathcal{R}_L^{'}$ is defined to be the part of
$\mathcal{R}_L$ not overlapping with $\mathcal{R}_T$.
Thus, the simultaneous fits are performed to 9
independent samples for $m_{A'}\le5.5$~GeV, and 4 independent samples
for $5.5<m_{A'}\le 8.0$~GeV
(missing mass spectra for all datasets are shown
in \cite{EPAPS}).

For the fits to the $\mathcal{R}_B$ regions, we
fix the number of signal events to zero, and determine the parameters of
the background PDFs. In the fits to the $\mathcal{R}_T$ and
$\mathcal{R}_L^{'}$ regions, we fix the background PDF
shape, and vary the number of background events $N_B$, the number of
peaking background events $e^+e^-\to\gamma\gamma$ (for
$m_{A'}\le5.5$~GeV), and the $A'$ mixing strength $\varepsilon^2$. The
numbers of signal and background events are constrained:
$\varepsilon^2\ge 0$ and $N_B>0$. 

The signal PDF is described by a Crystal Ball~\cite{ref:CBshape}
function centered around the expected value of
$M_X^2=m_{A'}^2$. We determine the PDF as a function of
$m_{A'}$ using high-statistics simulated samples of signal
events, and 
we correct it for the difference
between the photon energy resolution function in
data and simulation using a high-statistics
$e^+e^-\to\gamma\gamma$ sample in which one of the photons converts to
an $e^+e^-$ pair in the detector material~\cite{ref:bad2330}. 
The resolution for signal events decreases monotonically from
$\sigma(M_X^2)=1.5~\mathrm{GeV}^2$ for $m_{A'}\approx 0$ to 
$\sigma(M_X^2)=0.7~\mathrm{GeV}^2$ for $m_{A'}=8$~GeV. The background
PDF has two components: a peaking background from \gamgam\ events,
described by a Crystal Ball function, and a smooth function of $M_X^2$
dominated by the $e^+e^-\to\gamma e^+e^-$ (second order polynomial for
$m_{A'}\le 5.5$ and a sum of exponentiated polynomials for
$5.5<m_{A'}\le 8.0$~GeV).

The signal selection efficiency  varies slowly
as a function of $m_{A'}$ between 2.4-3.1\% ($\mathcal{R}_T$
selection for $m_{A'}\le 5.5$~GeV),
3.4-3.8\% ($\mathcal{R}_L^{'}$ for $m_{A'}\le 5.5$~GeV),
and $2.0-0.2\%$ ($\mathcal{R}_L$ selection for $5.5<m_{A'}\le 8.0$~GeV).

%
%
The largest systematic uncertainties in the signal yield are from the
shape of the signal and background PDFs, and the uncertainties in the
efficiency of signal and trigger selections. We determine the
uncertainty in the signal PDF by comparing the data
and simulated distributions of \gamgam\ events. We correct for the  
small observed differences, and use half of the correction as an estimate of
the systematic uncertainty.
We measure the trigger selection efficiency using
single-photon \gamgam\ and \bhabha\ events that are selected from a sample of
unbiased randomly accepted triggers. We find good agreement with
the simulation estimates of the trigger efficiency, within the
systematic uncertainty of $0.4\%$.
We compare the input BDT observables in simulation and in a sample of
the single-photon data events, counting the difference as a systematic
uncertainty of the signal selection efficiency. 
The total multiplicative error on the signal cross
section is $5\%$, and is small compared to the statistical
uncertainty.  
\begin{figure}[tb]
\bc
\includegraphics[width=0.49\textwidth]{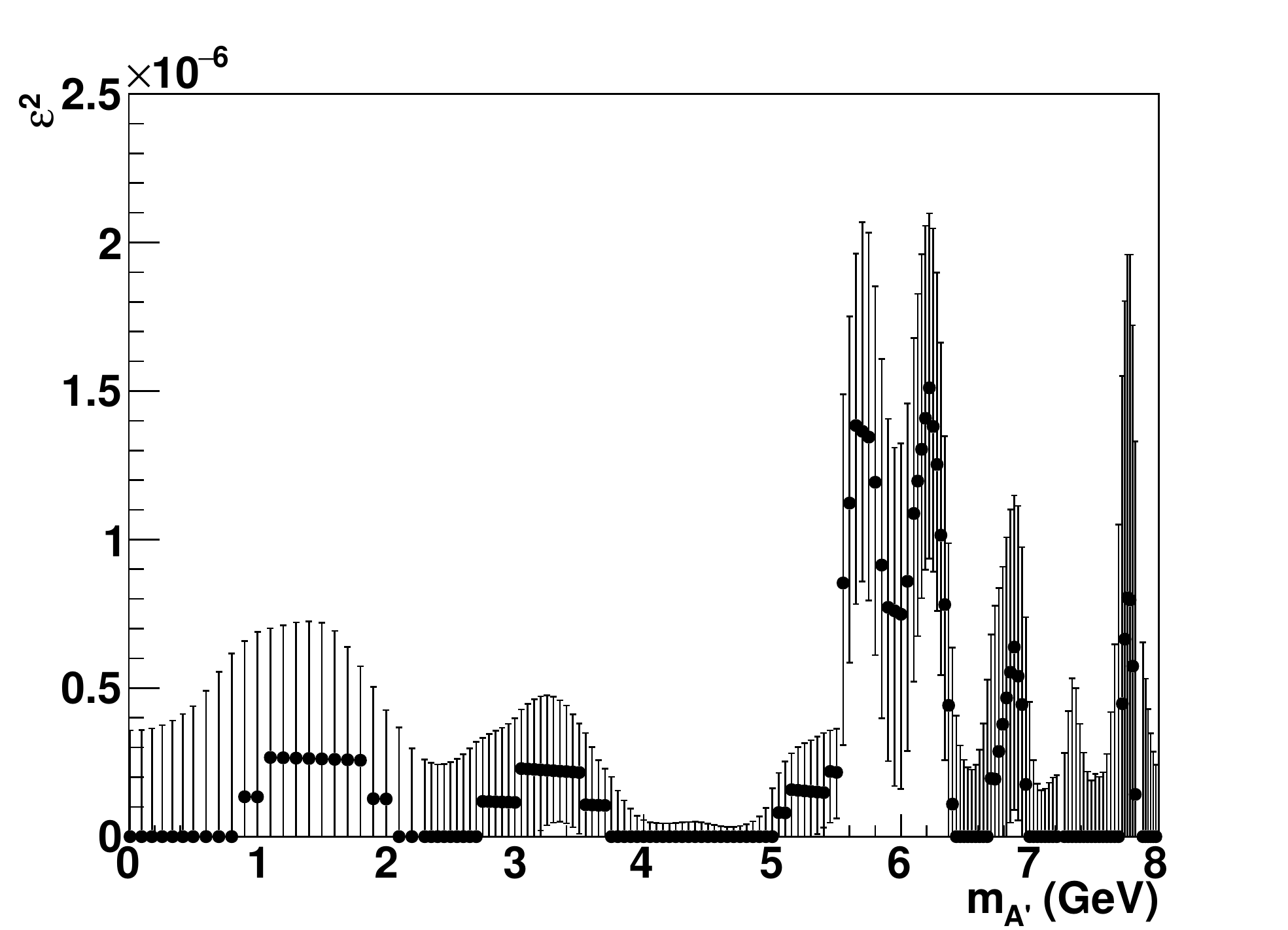}
\ec
\caption{Measured maximum-likelihood values of the $A'$ mixing strength squared
  $\varepsilon^2$ as a function of the mass $m_{A'}$.} 
\label{fig:point}
\end{figure}

\begin{figure}[tb]
\bc
\includegraphics[width=0.49\textwidth]{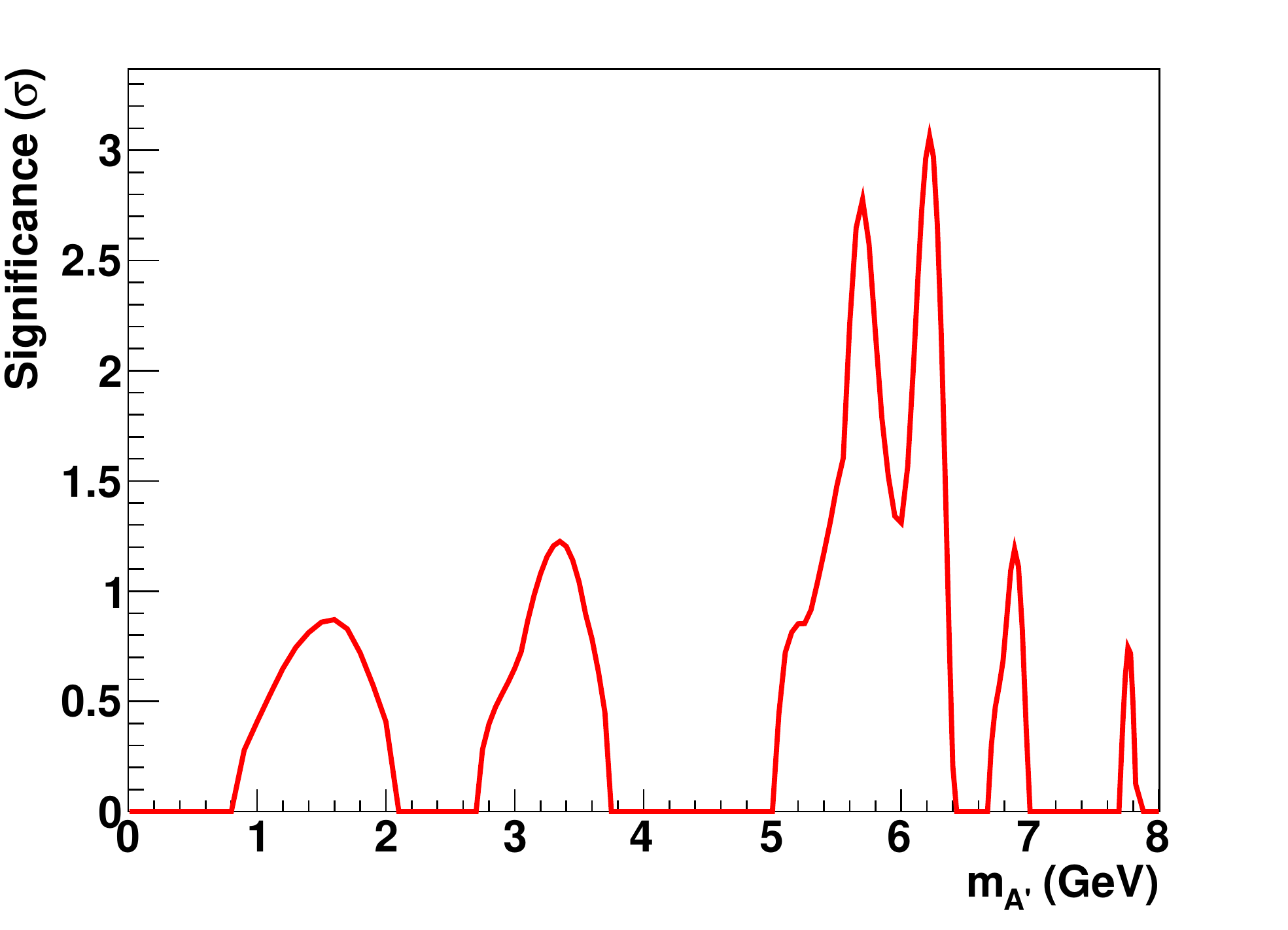}
\ec
\caption{Signal significance $\mathcal{S}$ as a
  function of the mass $m_{A'}$.} 
\label{fig:sign}
\end{figure}

\begin{figure}[tb]
\bc
\includegraphics[width=0.49\textwidth]{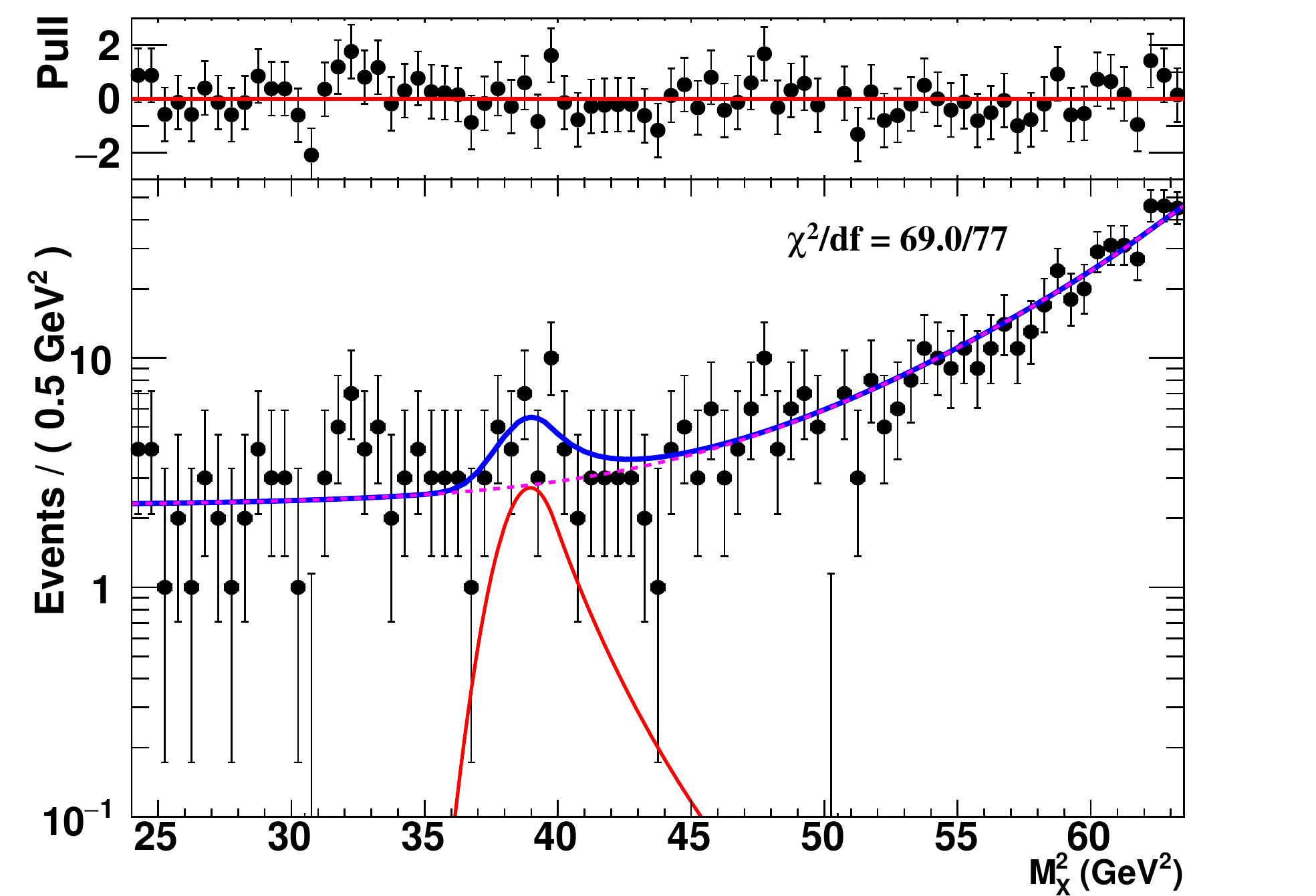}
\ec
\caption{Bottom: signal fit for $m_{A'}=6.21$~GeV to a combination of
  \Y2S\ and \Y3S\ datasets, shown for illustration purposes. 
  The signal peak (red) corresponds to the local significance
  $\mathcal{S}=3.1$ (global significance of $2.6\sigma$). Blue solid
  line shows the full PDF, while the magenta dashed line corresponds
  to the background contribution. Top: distribution of the normalized
  fit residuals (pulls).} 
\label{fig:fit621}
\end{figure}

Figure~\ref{fig:point} shows the maximum-likelihood estimators of the
$A'$ mixing strength $\varepsilon^2$ for the 166 $m_{A'}$ hypotheses. 
The values of ``local'' significance of observation 
$\mathcal{S}\equiv \sqrt{2\ln(L_{\max}/L_0)}$,
where $L_{\max}$ is the maximum value of the likelihood, and $L_0$ is the
value of the likelihood with the signal yield fixed to zero, are shown
in Fig.~\ref{fig:sign}. The
most significant deviation of $\epsilon^2$ from zero occurs at
$m_{A'}=6.21$~GeV and corresponds to $\mathcal{S}=3.1$. Parametrized
simulations determine
that the probability to find such 
a deviation in any of the 166 $m_{A'}$ points in the absence of any
signal is $\approx 1\%$,
corresponding to a ``global'' significance of $2.6\sigma$. 
A representative fit for $m_{A'}=6.21$~GeV is shown in
Fig.~\ref{fig:fit621}.


\begin{figure}[tb]
\bc
\includegraphics[width=0.49\textwidth]{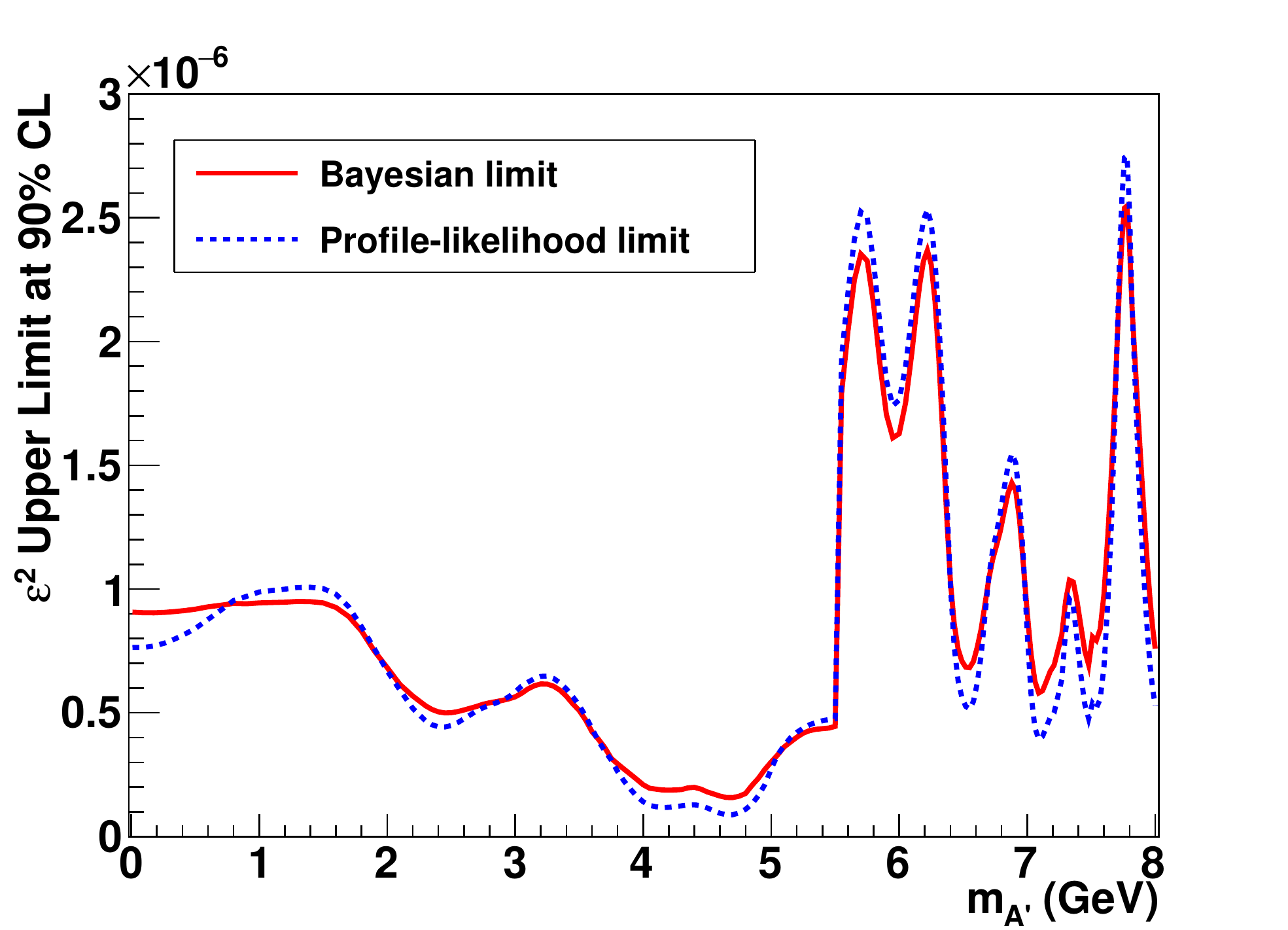}
\ec
\caption{Upper limits at 90\% CL on $A'$ mixing strength squared
  $\varepsilon^2$ as a 
  function of $m_{A'}$. Shown are the Bayesian limit computed with a
  uniform prior for $\varepsilon^2>0$ (solid red line) and the
  profile-likelihood limit (blue dashed line).} 
\label{fig:limit}
\end{figure}

\begin{figure}[tb]
\bc
\includegraphics[width=0.49\textwidth]{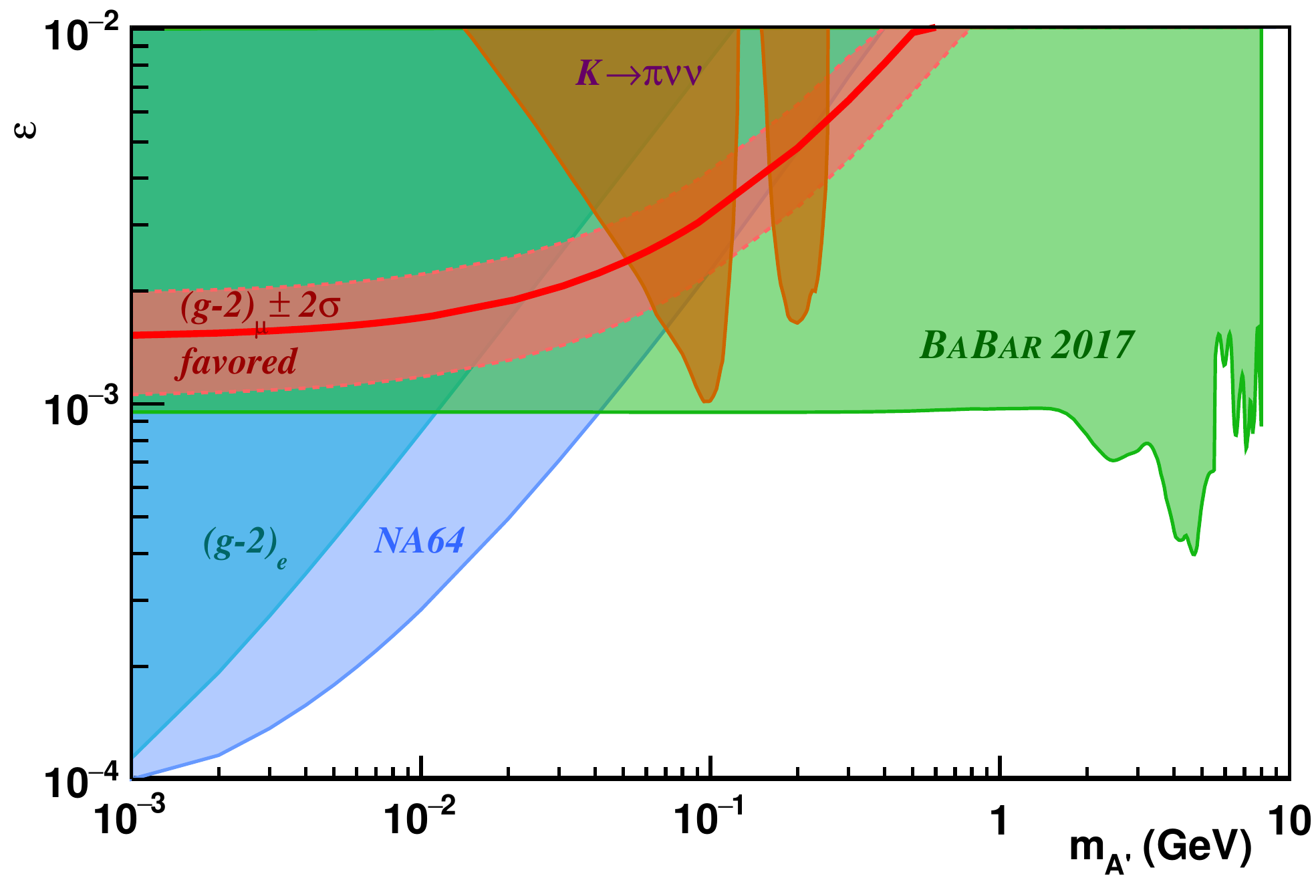}
\ec
\caption{Regions of the $A'$ parameter space ($\varepsilon$ vs $m_{A'}$)
  excluded by this work (green area) compared to the previous
  constraints~\cite{Essig:2013lka,E787,E949,NA64} as well as the
  region preferred by the 
  $(g-2)_\mu$  anomaly~\cite{ref:g-2}.}  
\label{fig:constraints}
\end{figure}

The 90\% confidence level (CL) upper limits on 
$\varepsilon^2$ as a function 
of $m_{A'}$ are
shown in Fig.~\ref{fig:limit}. We compute both the Bayesian limits with
a uniform prior for $\varepsilon^2>0$ and the frequentist profile-likelihood
limits~\cite{ref:Rolke}. Figure~\ref{fig:constraints} compares our results to
other limits on $\varepsilon$ in channels where $A'$ is allowed to decay
invisibly, as well as to the region of parameter space consistent with the
$(g-2)_\mu$ anomaly~\cite{ref:g-2}. At each value of $m_{A'}$ we
compute a limit on $\varepsilon$ as a square root of the Bayesian limit
on $\varepsilon^2$ from Fig.~\ref{fig:limit}.
Our data rules out the dark-photon
coupling as the explanation for the $(g-2)_\mu$ anomaly. Our limits
place stringent constraints on dark-sector models over a broad range
of parameter space, and represent a significant improvement over
previously available results. 

%

We are grateful for the excellent luminosity and machine conditions
provided by our \pep2\ colleagues, 
and for the substantial dedicated effort from
the computing organizations that support \babar.
The collaborating institutions wish to thank 
SLAC for its support and kind hospitality. 
This work is supported by the
US Department of Energy
and National Science Foundation, the
Natural Sciences and Engineering Research Council (Canada),
the Commissariat \`a l'Energie Atomique and
Institut National de Physique Nucl\'eaire et de Physique des Particules
(France), the
Bundesministerium f\"ur Bildung und Forschung and
Deutsche Forschungsgemeinschaft
(Germany), the
Istituto Nazionale di Fisica Nucleare (Italy),
the Foundation for Fundamental Research on Matter (The Netherlands),
the Research Council of Norway, the
Ministry of Education and Science of the Russian Federation, 
Ministerio de Econom\'{\i}a y Competitividad (Spain), the
Science and Technology Facilities Council (United Kingdom),
and the Binational Science Foundation (U.S.-Israel).
Individuals have received support from 
the Marie-Curie IEF program (European Union) and the A. P. Sloan Foundation (USA).


We wish to acknowledge Adrian Down, Zachary Judkins, and Jesse Reiss
for initiating the study of the physics opportunities with the single
photon triggers in \babar, Rouven Essig for stimulating
discussions and for providing data for Fig.~\ref{fig:constraints}, and
Farinaldo Queiroz for correcting a typo in Fig.~\ref{fig:constraints}.

\onecolumngrid
\newpage

\section{EPAPS Material}

The following includes supplementary material for the Electronic
Physics Auxiliary Publication Service. 

\begin{figure}[h!]
\begin{center}
\subfigure[]{\includegraphics[width=0.45\textwidth]{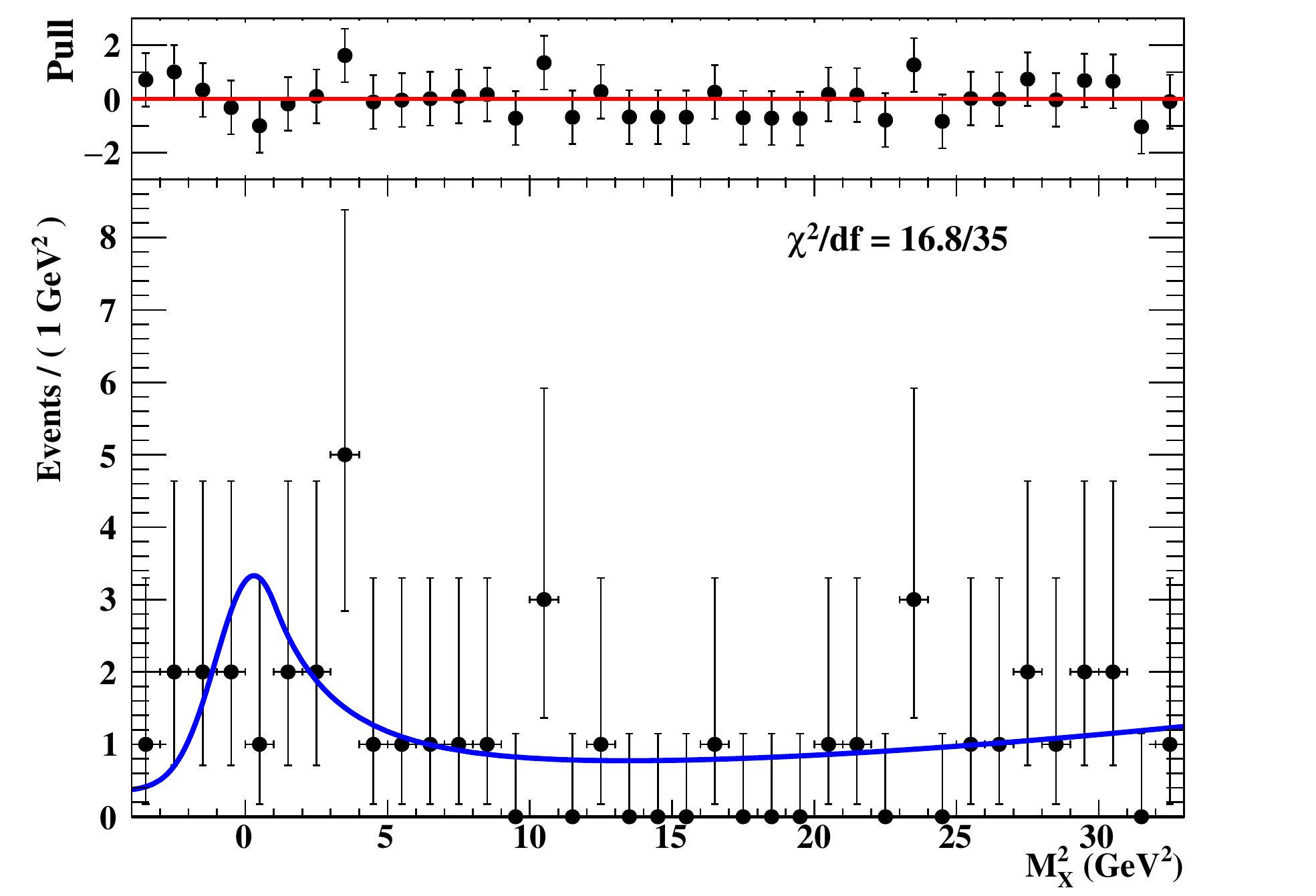}}
\subfigure[]{\includegraphics[width=0.45\textwidth]{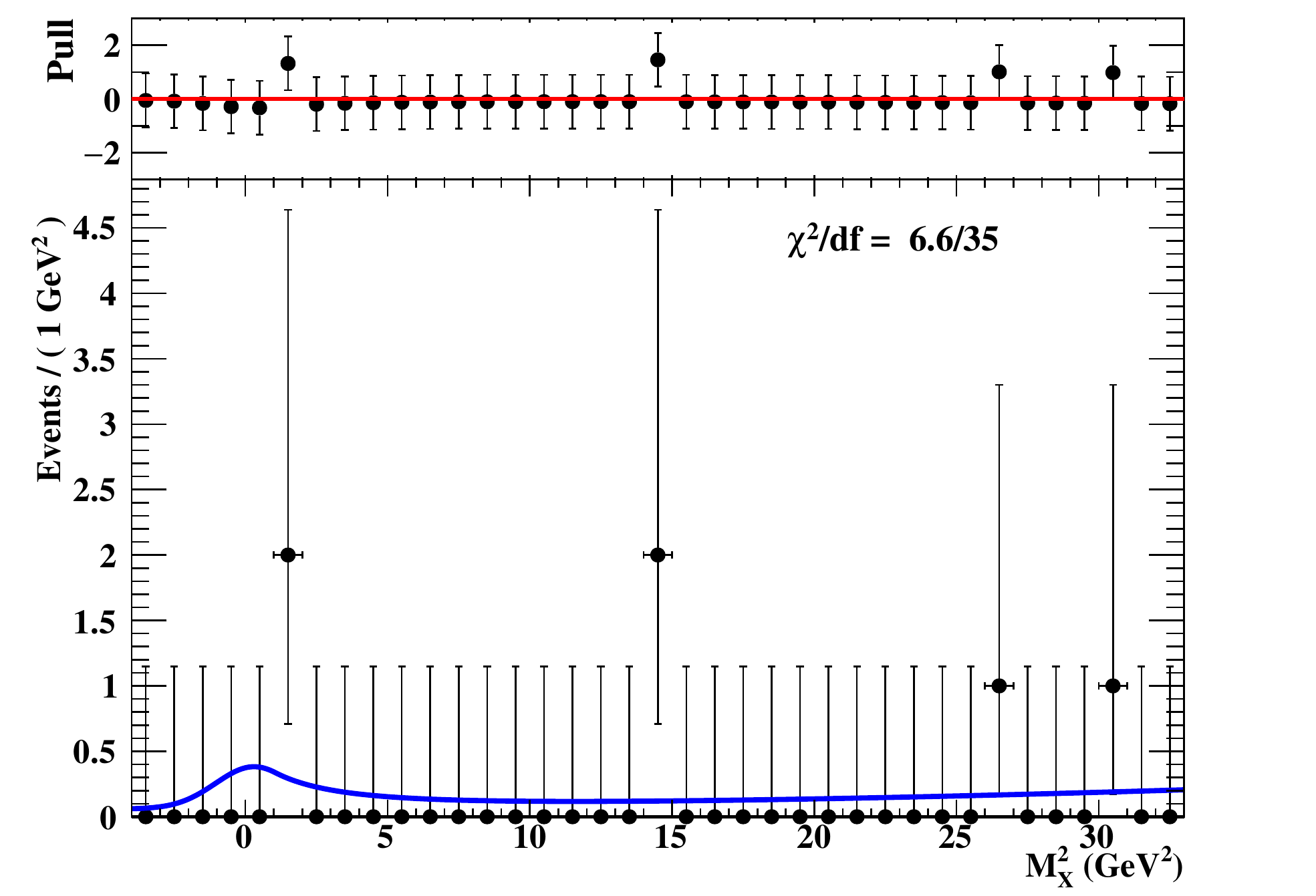}}
\subfigure[]{\includegraphics[width=0.45\textwidth]{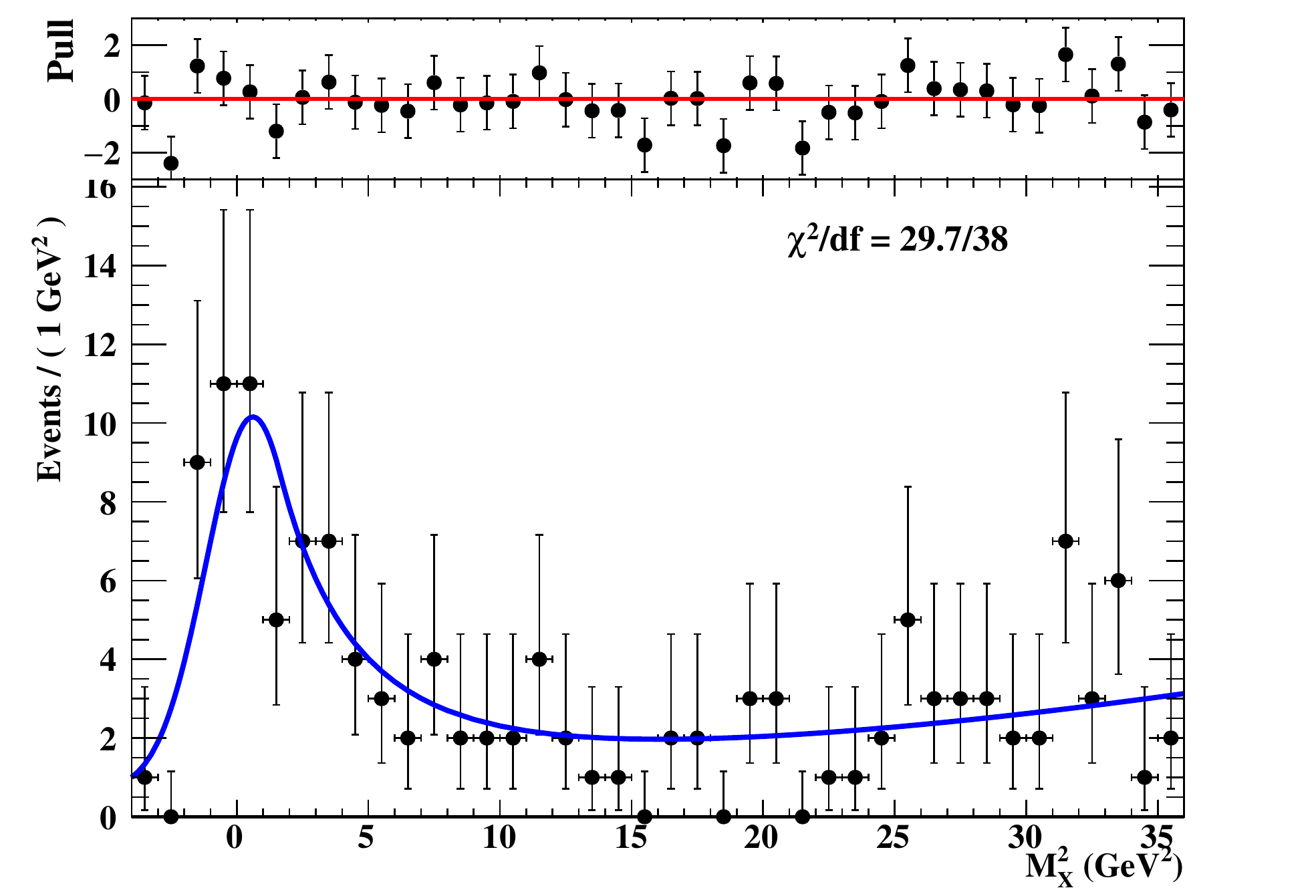}}
\subfigure[]{\includegraphics[width=0.45\textwidth]{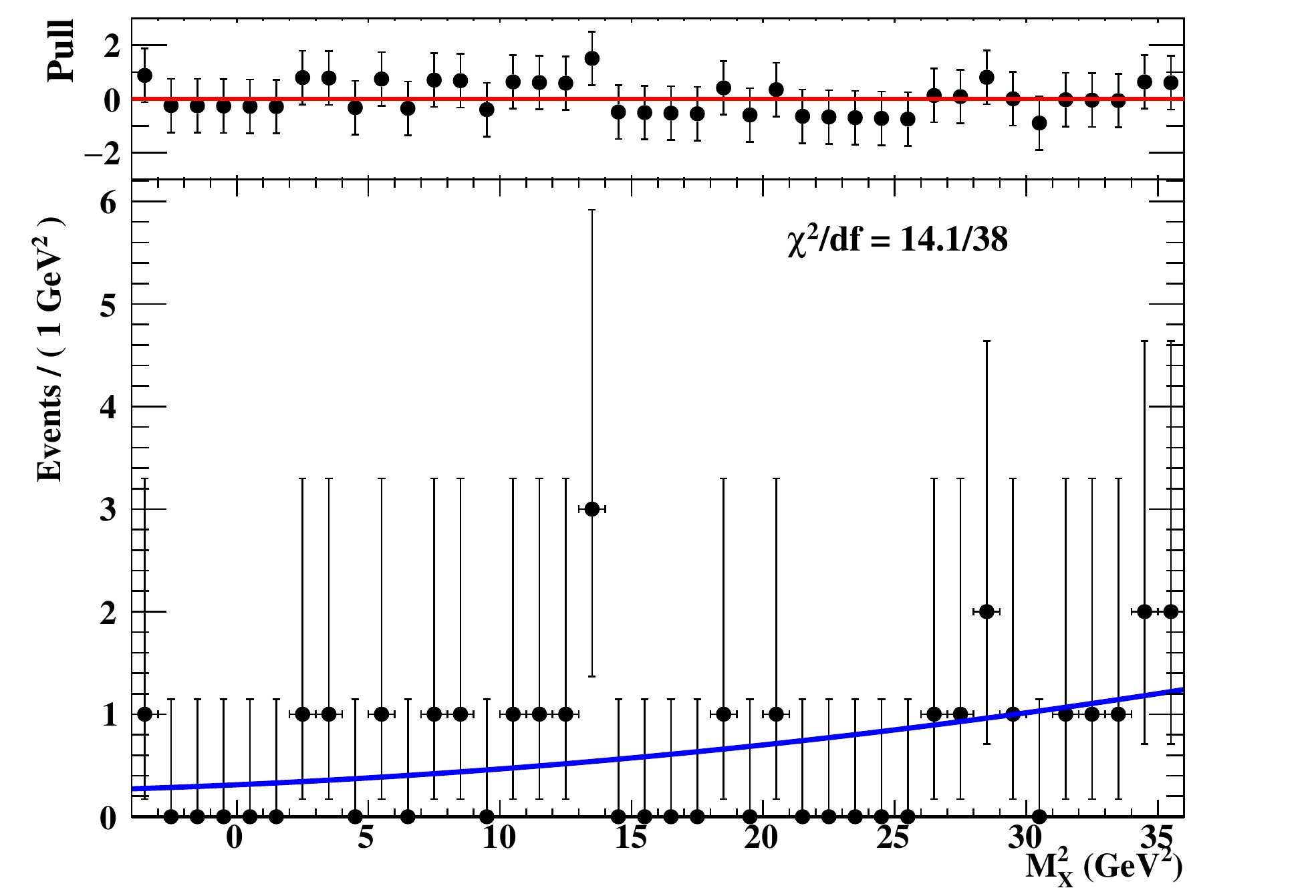}}
\subfigure[]{\includegraphics[width=0.45\textwidth]{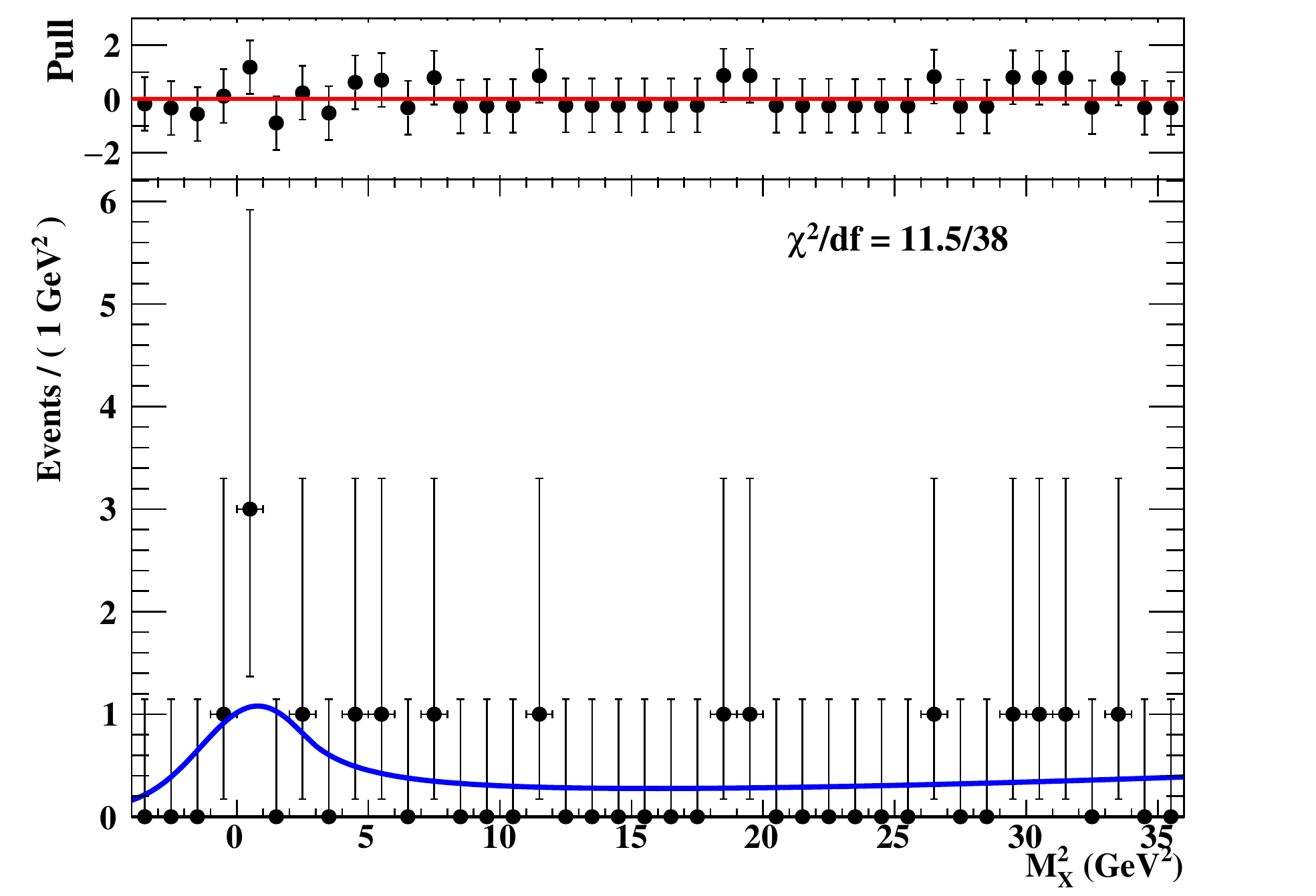}}
\subfigure[]{\includegraphics[width=0.45\textwidth]{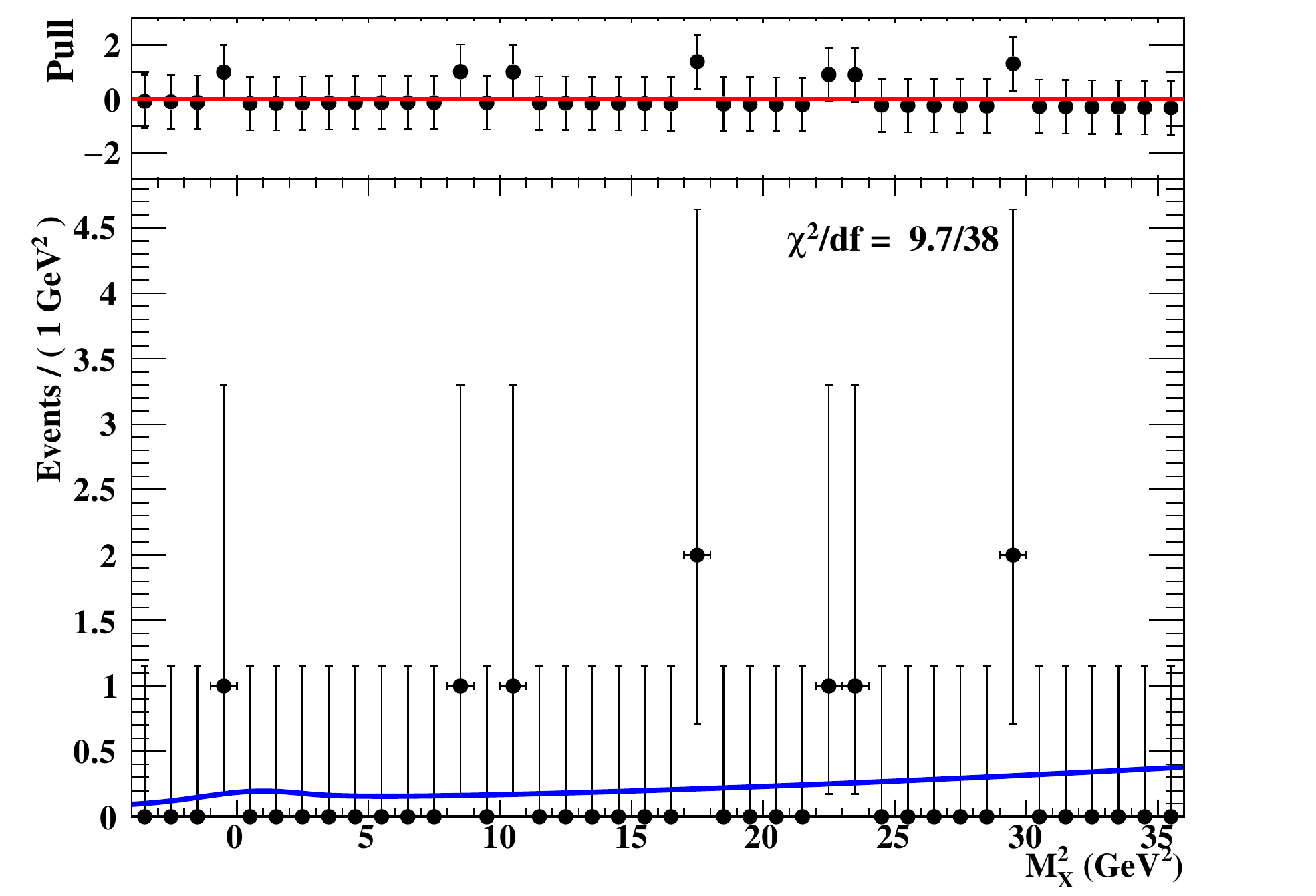}}
\end{center}
\caption{Distributions of the missing mass squared $M_X^2$ in the
  ``lowM'' data samples collected near (a,b) \Y2S, (c,d) \Y3S, and
  (e,f) \Y4S\ resonances. Data are selected with (a,c,e)
  $\mathcal{R}_L^{'}$ and (b,d,f) $\mathcal{R}_T$ selections. 
  The solid blue line represents the background-only fit with
  $\varepsilon^2\equiv0$. Normalized fit residuals are shown above
  each plot. 
}
\label{fig:spectraLowM}
\end{figure}
\begin{figure}[h!]
\begin{center}
\subfigure[]{\includegraphics[width=0.45\textwidth]{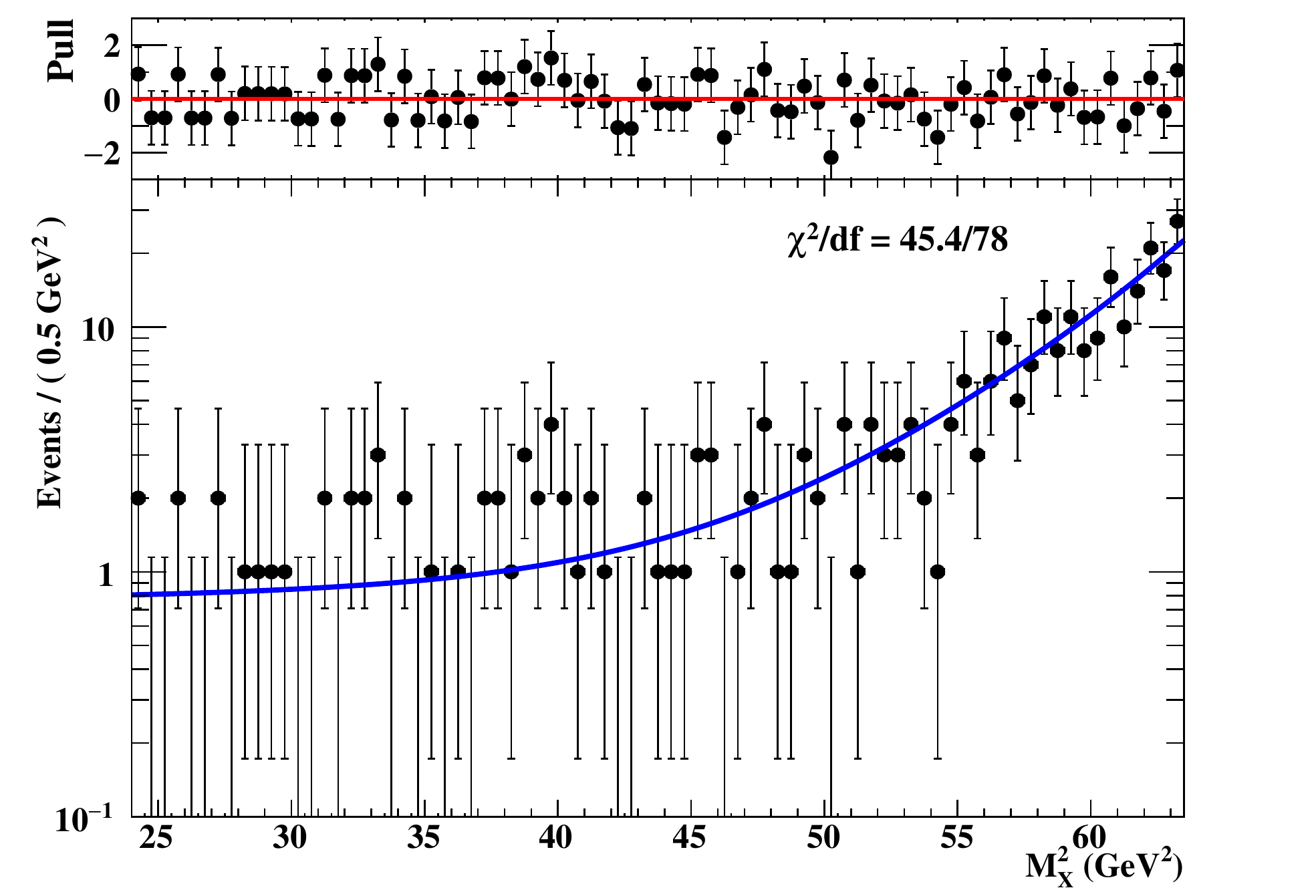}}
\subfigure[]{\includegraphics[width=0.45\textwidth]{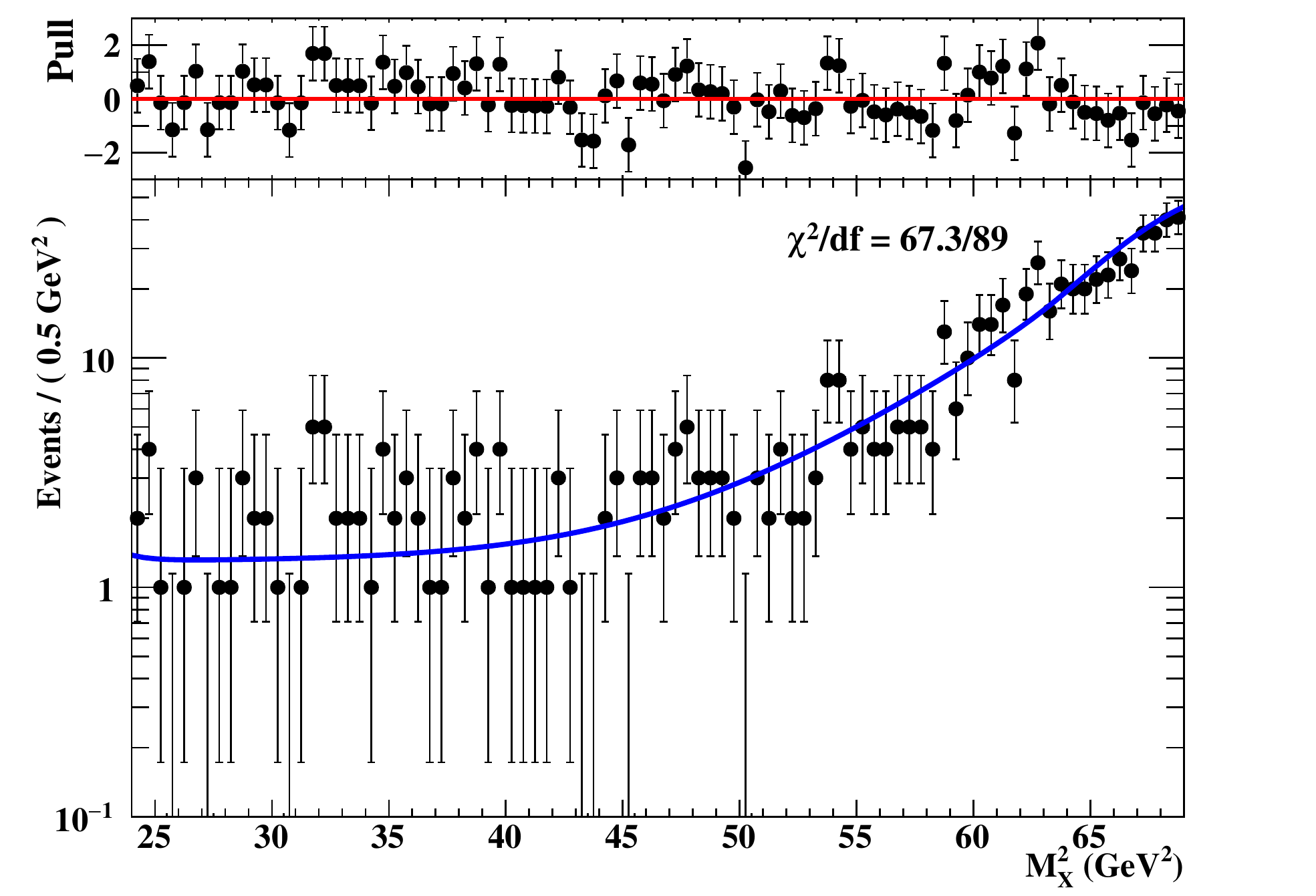}}
\end{center}
\caption{Distributions of the missing mass squared $M_X^2$ in the
  ``highM'' data samples collected near (a) \Y2S\ and (b)
  \Y3S\ resonances. 
  The solid blue line represents the background-only fit with
  $\varepsilon^2\equiv0$. Normalized fit residuals are shown above
  each plot. 
}
\label{fig:spectraHighM}
\end{figure}


\begin{thebibliography}{999}

\bibitem{ref:INTEGRAL}
P.~Jean \etal,
  Astron.\ Astrophys.\  {\bf 407}, L55 (2003); \\
  J.~Knodlseder \etal,
  Astron.\ Astrophys.\  {\bf 411}, L457 (2003).

\bibitem{ref:PAMELA}
O.~Adriani \etal\ [PAMELA Collaboration], Nature \textbf{458}, 607 (2009).  

\bibitem{ref:FERMI}
M.~Ackermann \etal\ [Fermi LAT Collaboration],
Phys. Rev. Lett. \textbf{108}, 011103 (2012).  

\bibitem{Berezhiani:2013dea} 
  Z.~Berezhiani, A.~D.~Dolgov and I.~I.~Tkachev,
  Eur.\ Phys.\ J.\ C {\bf 73}, 2620 (2013).


\bibitem{ref:g-2} 
G.~W.~Bennett \etal\ [Muon g-2 Collaboration],
  Phys.\ Rev.\ D {\bf 73}, 072003 (2006).

\bibitem{ref:Aprimerefs}
P. Fayet, Phys. Lett. B \textbf{95} 285 (1980),
Nucl. Phys. B \textbf{187}, 184 (1981);
B.~Holdom, Phys. Lett. B \textbf{166}, 196 (1986);
N.~Borodatchenkova, D.~Choudhury and M.~Drees,
Phys.\ Rev.\ Lett.\  {\bf 96}, 141802 (2006);
D.~P.~Finkbeiner and N.~Weiner, Phys. Rev. D \textbf{76}, 083519
(2007); 
M.~Pospelov, A.~Ritz, and M.~B.~Voloshin,
Phys. Lett. B \textbf{662}, 53 (2008);
N. Arkani-Hamed \etal, Phys. Rev. D \textbf{79}, 015014 (2009).

\bibitem{Essig:2013lka}
R.~Essig \etal, arXiv:1311.0029 [hep-ph], and references therein.  

\bibitem{BaBarDM}
J.~P.~Lees \etal\ [\babar\ Collaboration],
Phys.\ Rev.\ Lett.\  {\bf 113}, 201801 (2014).

\bibitem{KLOE}
D.~Babusci \etal\ [KLOE-2 Collaboration],
Phys.\ Lett.\ B {\bf 720}, 111 (2013);
Phys.\ Lett.\ B {\bf 736}, 459 (2014).

\bibitem{NA48}
J.~R.~Batley \etal\ [NA48/2 Collaboration],
Phys.\ Lett.\ B {\bf 746}, 178 (2015).

\bibitem{WASA}
P.~Adlarson \etal\ [WASA-at-COSY Collaboration],
Phys.\ Lett.\ B {\bf 726}, 187 (2013).

\bibitem{HADES} 
G.~Agakishiev \etal\ [HADES Collaboration],
Phys.\ Lett.\ B {\bf 731}, 265 (2014).

\bibitem{A1} 
H.~Merkel \etal\ [A1 Collaboration],
Phys.\ Rev.\ Lett.\  {\bf 112}, 221802 (2014).

\bibitem{APEX} 
S.~Abrahamyan \etal\ [APEX Collaboration],
Phys.\ Rev.\ Lett.\  {\bf 107}, 191804 (2011).


\bibitem{Blumlein:2013cua} 
J.~Bl\"umlein and J.~Brunner,
Phys.\ Lett.\ B {\bf 731}, 320 (2014).

\bibitem{Andreas:2012mt} 
S.~Andreas, C.~Niebuhr and A.~Ringwald,
Phys.\ Rev.\ D {\bf 86}, 095019 (2012).

\bibitem{Pospelov}
M.~Pospelov,
Phys. Rev. D \textbf{80}, 095002 (2009).

\bibitem{E787}
S.~Adler \etal\ [E787 Collaboration],
Phys.\ Rev.\ Lett.\  {\bf 88}, 041803 (2002).
  

\bibitem{E949}
A.~V.~Artamonov \etal\ [E949 Collaboration],
Phys. Rev. D \textbf{79}, 092004 (2009).

\bibitem{NA64}
D.~Banerjee {\it et al.} [NA64 Collaboration],
Phys.\ Rev.\ Lett.\ \textbf{118}, 011802 (2017).
 


\bibitem{Lumi} 
  J.~P.~Lees {\it et al.} [\babar\ Collaboration],
  Nucl.\ Instrum.\ Meth.\ A {\bf 726}, 203 (2013).

\bibitem{detector}
B.\ Aubert \etal\ [\babar\ Collaboration],
  Nucl.\ Instrum.\ Meth.\ A {\bf 479}, 1 (2002);
Nucl.\ Instrum.\ Meth.\ A {\bf 729}, 615 (2013).

\bibitem{geant}
S. Agostinelli \etal\ [{\sc GEANT4\/} Collaboration],
  Nucl.\ Instrum.\ Meth.\ A {\bf 506}, 250 (2003).

\bibitem{TMVA}
A. Hoecker \etal,
PoS ACAT 040 (2007), arXiv:physics/0703039.

\bibitem{ref:LAT}
A.\ Drescher \etal [ARGUS Collaboration], 
\nima{237},  464 (1985).

\bibitem{EPAPS}
Additional plots are available through EPAPS Document No. E-PRLTAO-XX-XXXXX.
For more information on EPAPS, see http://www.aip.org/pubservs/epaps.html. 

\bibitem{ref:CBshape}
M.~J.~Oreglia, Ph.D. Thesis, report SLAC-236 (1980), Appendix D;
J.~E.~Gaiser, Ph.D. Thesis, report SLAC-255 (1982), Appendix F;
T.~Skwarnicki, Ph.D. Thesis, report DESY F31-86-02(1986), Appendix E.

\bibitem{ref:bad2330} 
P.~del Amo Sanchez \etal\ [\babar\ Collaboration],
  Phys.\ Rev.\ Lett.\  {\bf 107}, 021804 (2011).

\bibitem{ref:Rolke} 
W.~A.~Rolke, A.~M.~Lopez and J.~Conrad,
Nucl.\ Instrum.\ Meth.\ A {\bf 551}, 493 (2005).


\end{thebibliography}
\end{document}